\documentclass[preprint,showpacs,keywords,preprintnumbers,amsmath,amssymb]{revtex4}

\usepackage{graphicx}
\usepackage{dcolumn}
\usepackage{bm}

\begin{document}


\title{ Conformal Scaling Gauge Symmetry and Inflationary Universe }
\author{Yue-Liang Wu }
 \email{ylwu@itp.ac.cn}
 \affiliation{Institute of Theoretical Physics, Chinese Academy of Sciences,
 Beijing 100080, China  }
 %
\date{\today}
\begin{abstract}
Considering the conformal scaling gauge symmetry as a fundamental
symmetry of nature in the presence of gravity, a scalar field is
required and used to describe the scale behavior of universe. In
order for the scalar field to be a physical field, a gauge field
is necessary to be introduced. A gauge invariant potential action
is constructed by adopting the scalar field and a real Wilson-like
line element of the gauge field. Of particular, the conformal
scaling gauge symmetry can be broken down explicitly via fixing
gauge to match the Einstein-Hilbert action of gravity. As a
nontrivial background field solution of pure gauge has a minimal
energy in gauge interactions, the evolution of universe is then
dominated at earlier time by the potential energy of background
field characterized by a scalar field. Since the background field
of pure gauge leads to an exponential potential model of a scalar
field, the universe is driven by a power-law inflation with the
scale factor $a(t) \sim t^p$. The power-law index $p$ is
determined by a basic gauge fixing parameter $g_F$ via $p = 16\pi
g_F^2[1 + 3/(4\pi g_F^2) ]$. For the gauge fixing scale being the
Planck mass, we are led to a predictive model with $g_F=1$ and
$p\simeq 62$.
\end{abstract}

\maketitle

Recent developments in cosmology\cite{COS} indicate that the
astrophysics reaches an epoch for exploring fundamental theory
inaccessible to particle accelerators. Especially due to more
accurate astrophysical data from Wilkinson Microwave Anisotropy
Probe (WMAP)\cite{CC} and other planned future cosmological
observations, it enables one to test various cosmological models.
One of the basic ideas in modern cosmology is the
inflation\cite{IF1,IF2,IF3,IF4,COS} which postulates that the
universe at earlier times in its history was dominated by the
potential or vacuum energy. The developments of inflationary
cosmology involve various scalar fields and different types of
potentials, while all proposed models attend to solve the
difficult cosmological problems met in the standard Big Bang
cosmology, such as the homogeneity, isotropy, horizon, flatness,
structure formation, monopole {\it etc.}. It is intriguing that
the constituents in today's universe are also dominated by the
potential or vacuum energy, i.e., the so-called dark energy, about
$73\%$ by weight. The remaining party of universe is mainly filled
by the so-called dark matter, about $23\%$ by weight, while the
density of ordinary baryonic matter known from the standard model
in particle physics is less than $4\%$ by weight. Namely all
structure in the universe, from small scale to large scale, may be
formed via a common origin based on the potential or vacuum
energy. Therefore either the inflation happened in the small scale
of universe at earlier times or the accelerated expansion observed
in the large scale of today's universe can be driven by dynamical
scalar fields or homogenous dark energy. Obviously, those scalar
fields must be beyond the standard model in particle physics. It
then rises some basic issues, such as: whether the introduced
scalar fields are the fundamental fields; what are the deep
physical reasons for introducing scalar fields; how the scalar
fields interact with ordinary matter fields; which kind of
potential form or vacuum structure is the true choice of nature;
why the small and large scale behavior of universe is so
analogous. In general, different motivations and ideas lead to
different inflationary models. Nevertheless, recent progresses in
more accurate observational cosmology provide a possibility to
have an experimental verification for various inflationary models.
It is expected that more precise cosmological observations will
guide us to a more fundamental theory.

 In here the large and small scales of universe are characterized by the
corresponding low and high mass energy scales $M_{Hl}$ and
$\bar{M}_{Pl}$ respectively. $M_{Hl}\equiv H_0 = 2.0\times
10^{-33}$ eV is the Hubble constant and $\bar{M}_{Pl} =
M_{Pl}/\sqrt{8\pi} = 2.44\times 10^{18}$ GeV = $ 2.44\times
10^{27}$ eV is the reduced Planck mass ($M_{Pl} = 1.22\times
10^{19}$ GeV). A mysterious observation is that the cosmological
constant $\Lambda$ or the homogeneous dark energy is found at the
present epoch to be at the energy scale given by the relation
\begin{eqnarray}
\frac{\bar{M}_{Pl}}{\Lambda} \simeq \frac{\Lambda}{M_{Hl}}, \quad
\mbox{i.e.} \quad  \Lambda \simeq \sqrt{\bar{M}_{Pl}M_{Hl} } =
2.21 \times 10^{-3} eV
\end{eqnarray}
In general, the introduction of mass scale in a four dimensional
space-time theory will spoil a conformal scaling global symmetry
in the absence of gravity or destroy a conformal scaling gauge (or
local) symmetry in the presence of gravity. It is noticed that a
conformal scaling gauge symmetry, which differs from a conformal
scaling global symmetry, implies the existence of a fundamental
mass scale as one can always make a conformal scaling gauge
transformation to fix the gauge so as to match the
Einstein-Hilbert action of gravity. Namely, the conformal scaling
gauge symmetry can be made to be broken down explicitly by an
appropriate fixing gauge, which distinguishes from the spontaneous
symmetry breaking of unitary gauge symmetries.

From the special feature of conformal scaling gauge symmetry, we
are going to study in this note a theory based on an assumption
that the conformal scaling gauge symmetry is a fundamental
symmetry in the presence of gravity. It is then known that a
scalar field $\Phi(x)$ is necessary for constructing a conformal
scaling gauge invariant action of gravity. While such a scalar
field is not a physical one due to a wrong sign of its kinetic
term. To make the scalar field be a dynamically physical field, it
is seen that a gauge field is necessary to be introduced for
obtaining a gauge invariant kinetic term for the scalar field. We
then show that a conformal scaling gauge invariant potential
action can in general be constructed by the scalar field and a
real Wilson-like line element of gauge field. As a nontrivial
background field solution of pure gauge has a minimal energy in
gauge interactions, the evolution of universe is dominated by the
potential energy of the pure gauge background field which is
characterized by a scalar field. As a consequence, we naturally
arrive at an exponential potential model of the scalar field after
fixing the gauge to match the Einstein-Hilbert action of gravity.
It will be seen that such a model will lead to an inflationary
universe\cite{IF1,IF2,IF3} via a chaotic type inflation\cite{IF4}.
As the exponential potential model has an exact solution with a
power-law inflation\cite{PL} for the scale factor,i.e., $a(t) \sim
t^p$, its cosmological effects are solely characterized by the
power-law index $p$. It will be shown that the power-law index in
such an inflationary theory is determined by a basic gauge fixing
parameter $g_F$ via $p = 16\pi g_F^2 [ 1 + 3/(4\pi g_F^2) ]$. When
taking the Planck mass to be the gauge fixing scale of the scalar
field, we have $g_F = 1$ and $p\simeq 62$. The resulting
cosmological effects in such an inflationary theory will be
discussed.

The conformal scaling gauge invariant action for gravity requires
the introduction of a fundamental scalar field $\Phi(x)$. It has
the following known form
\begin{eqnarray}
S_G = \int d^4 x \sqrt{-g}\ \kappa_0^2 \{ -\frac{1}{2}g^{\mu\nu}
\partial_{\mu}\Phi \partial_{\nu} \Phi - \frac{1}{12} \Phi^2 R \}
\end{eqnarray}
with $\kappa_0$ a constant. Where $g_{\mu\nu}$ is a metric which
has the signature (+ - - -). This action is invariant under the
conformal scaling gauge transformation
\begin{eqnarray}
\Phi(x) \rightarrow \xi(x) \Phi(x), \qquad g_{\mu\nu}(x)
\rightarrow \xi^{-1}(x) g_{\mu\nu}(x)
\end{eqnarray}

Note that the above kinetic term of the scalar field has a wrong
sign for the scalar field to be as a physical one. To construct an
invariant kinetic term of the scalar field with a right sign, it
is necessary to introduce a gauge field $A_{\mu}(x)$ with
transformation property
\begin{eqnarray}
A_{\mu}(x) \rightarrow A_{\mu}(x) - \xi^{-1}(x)\partial_{\mu}
\xi(x)
\end{eqnarray}
Thus the invariant kinetic action can be written down as follows
\begin{eqnarray}
S_K = \int d^4 x \sqrt{-g} \{ \frac{1}{2} \kappa^2 g^{\mu\nu}
D_{\mu}\Phi D_{\nu} \Phi  -\frac{1}{4g^2} F_{\mu\nu}F^{\mu\nu} \}
\end{eqnarray}
with $\kappa$ a constant satisfying the normalization condition of
kinetic term
\begin{eqnarray}
 \kappa^2 - \kappa_0^2 = 1
\end{eqnarray}
Where the covariant derivation and the gauge field strength are
defined as
\begin{eqnarray}
 & & D_{\mu}\Phi(x) = (\partial_{\mu} - A_{\mu} ) \Phi(x) \\
 & & F_{\mu\nu} = \partial_{\mu}A_{\nu} - \partial_{\nu}A_{\mu}
\end{eqnarray}
Before proceeding, we would like to point out that the conformal
scaling gauge field $A_{\mu}$ do not directly couple to fermionic
fields $\psi_i(x)$. This is because the fermionic action
\begin{eqnarray}
 S_F = \int d^4 x \sqrt{-g} \{ \bar{\psi}_i \gamma^a e^{\mu}_a
 \left(i\partial_{\mu} + \omega_{\mu}^{bc}\Sigma_{bc} \right) \psi_i +
 H.c. \}
\end{eqnarray}
is invariant under the conformal scaling gauge transformation when
the fermion fields transform as $\psi_i \to \xi^{3/2}(x) \psi_i$.
There is also no Yukawa coupling terms $\bar{\psi}_i \psi_i \Phi $
for the ordinary fermion fields in the standard model of particle
physics. This is because the fermion fields in the standard model
are chiral with gauge symmetry $SU(2)_L \times U(1)_Y \times
SU(3)_c$. Nevertheless, the scalar field $\Phi$ may have
interaction with singlet fermions beyond the standard model of
particle physics, such as the right-handed Majorana neutrinos with
$y_N \bar{N}_R N_R^c \Phi $. In general, the scalar field $\Phi$
and gauge field $A_{\mu}$ can interact with the ordinary matter
fields only via a gravitational interaction. It naturally avoids
an observable strongly interacting fifth force in the conformal
scaling gauge theory.

We now consider a real Wilson-like line element associated the
element in the conformal scaling gauge group for the path $C_P$
going from $x_P$ to $x$ with the gauge field $A_{\mu} (x)$
 \begin{eqnarray}
  G_P(x, C_P; A) = P exp \int_{x_P}^x dz^{\mu} A_{\mu} \
 \end{eqnarray}
with $P$ the path-ordering operation. It transforms under the
conformal gauge transformation as follows
 \begin{eqnarray}
 G_P(x, C_P; A) \rightarrow \xi(x) G_P(x, C_P; A)\xi^{-1}(x_P)
 \end{eqnarray}
Here $x_P$ is a point at a small scale of Planck length
$l_p=1/M_{Pl}$. With the above Wilson-like line element, we can in
general construct the following gauge invariant potential
\begin{eqnarray}
S_V & \equiv & -\int d^4 x \sqrt{-g}\  V(A_{\mu}, \Phi) \\
& = & - \int d^4 x \sqrt{-g} \sum_{n=1}^{4} \lambda_{n}
\Phi^{4-n}(x)[G_P(x, C_P; A) \Phi(x_P)]^{n}
\end{eqnarray}
with $\lambda_{n}$ being coupling constants.

  It is known that the pure gauge solution has a minimal energy in
  gauge interactions. Taking the pure gauge as a nontrivial background field solution of
gauge interactions and supposing that such a solution is dominated
at earlier time of universe, thus the evolution of universe is
described by the background field of pure gauge. For the Abelian
conformal scaling gauge symmetry, the pure gauge background field
is simply characterized by a scalar field $\chi(x)$.  In general
we may rewrite the gauge field into two parts
\begin{eqnarray}
A_{\mu}= \partial_{\mu} \ln \chi  + g\ a_{\mu} =
\chi^{-1}\partial_{\mu} \chi  + g\ a_{\mu}
\end{eqnarray}
where $g$ is the coupling constant and $a_{\mu}$ represents the
quantum fluctuation of gauge field in the background field of pure
gauge. The conformal scaling gauge transformation property for the
scalar field is $\chi(x) \rightarrow \xi(x) \chi(x)$, while the
quantum field $a_{\mu}$ is unchanged under the transformation. For
the pure gauge part, the Wilson-like line element $G_{P}(x, C_{P};
\partial_{\mu} \ln \chi)$ becomes independent of the gauge field
and gets the following simple form
\begin{eqnarray}
 G_{P}(x, C_{P}; \partial_{\mu} \ln \chi) =  \frac{\chi(x)}{\chi(x_P) }
 \end{eqnarray}
 We assume that around the point $x_P$ the scalar field $\chi(x)$
and $\Phi(x)$ run into a conformal scaling gauge invariant fixed
point so that their ratio is fixed to be a constant
\begin{eqnarray}
 \Phi(x_P)/\chi(x_P) = c_0
\end{eqnarray}
Thus an invariant action with a pure gauge background field can
simply be expressed as follows
\begin{eqnarray}
& & S = S_G +  S_K  + S_V  \nonumber \\
& & \to \int d^4 x \sqrt{-g} \{  \kappa_0 [ -\frac{1}{2}g^{\mu\nu}
\partial_{\mu}\Phi \partial_{\nu} \Phi - \frac{1}{12} \Phi^2 R ] + \frac{1}{2} \kappa^2
g^{\mu\nu} D_{\mu}\Phi D_{\nu} \Phi - \sum_{n=1}^{4} \lambda_{n}
c_0^n \Phi^{4-n} \chi^n \}
 \nonumber
\end{eqnarray}

On the other hand, one can always make a conformal scaling gauge
transformation to fix the gauge, so that the conformal scaling
gauge symmetry becomes manifestly broken down in such a fixing
gauge. In the presence of gravity, we can choose the most
convenient gauge fixing condition in such a way that it leads to
the Einstein-Hilbert action for gravity. Regarding the Planck mass
is a fundamental energy scale that characterizes the scaling
behavior of universe, it is then natural to choose the following
gauge fixing condition
\begin{eqnarray}
 & & \Phi(x) \rightarrow \xi(x) \Phi(x) \equiv g_F M_{Pl}  \\
 & & \chi(x) \to \xi(x) \chi(x) \equiv \hat{\chi}(x) \\
 & & g_{\mu\nu}(x) \rightarrow \xi^{-2}(x) g_{\mu\nu}(x) \equiv
 \hat{g}_{\mu\nu}
\end{eqnarray}
where the gauge fixing parameter $g_F$ is a basic parameter
introduced for a general case. With the above gauge fixing
condition the action for gravity and pure gauge background field
is simplified to be
\begin{eqnarray}
S_0 = \int d^4 x \sqrt{-\hat{g}} \{ \frac{1}{2}  \hat{g}^{\mu\nu}
\partial_{\mu}\phi \partial_{\nu} \phi   - V(\phi) -
 \frac{1}{2} \bar{M}_{Pl}^2 R \}
\end{eqnarray}
with
\begin{eqnarray}
& &  V(\phi)  = \sum_{n=1}^{4}  V_{n} (\phi)
 \equiv \sum_{n=1}^{4} V_n^0 \ e ^{n\phi/M}, \qquad V_n^0 =
 \lambda_n g_F^{4-n} c_0^n M_{Pl}^4 \\
 & & \kappa g_F M_{Pl} \ln (\hat{\chi}(x)/M_{Pl}) = M \ln (\hat{\chi}(x)/M_{Pl})
 \equiv \phi(x), \qquad \phi(x_P) = 0
\end{eqnarray}
where we have ignored the quantum gauge field $a_{\mu}$, its
effects are assumed to be unimportant during the inflationary
epoch, especially for the case with a small gauge coupling
constant $g\ll 1$.

 Taking the above gauge fixing condition, together with
the normalization condition for the general relativity of Einstein
theory and the normalization condition for the kinetic term of
scalar field, the relevant three basic parameters can be expressed
in terms of one basic gauge fixing parameter $g_F$
\begin{eqnarray}
& & \kappa_0 = \frac{1}{g_F}\ \sqrt{\frac{3}{4\pi}}, \qquad \kappa
= \sqrt{1 +
\frac{3}{4\pi g_F^2}} \\
 & & M = \kappa g_F M_{Pl} = M_{Pl} \sqrt{g_F^2 +
\frac{3}{4\pi}}
\end{eqnarray}

Thus we arrive at an exponential potential model for the scalar
field $\phi$. It will be shown that the scalar field $\phi$ can
drive an inflationary universe at earlier time.
 The vacuum solution for the scalar field is
 obtained by solving the minimal condition for the scalar field.
 For an inflationary universe, the vacuum expectation values of the scalar field are
 in general time-dependent. Assuming that during that stage the universe
 is described in a good approximation by a
 Robertson-Walker metric
 \begin{eqnarray}
 \hat{g}_{\mu\nu}(x) \equiv g^{RW}_{\mu\nu} + \tilde{g}_{\mu\nu}(x)
 \simeq  g^{RW}_{\mu\nu}
 \end{eqnarray}
 Thus the time evolution of scalar field is governed by its equation of
 motion in the cosmological background. Namely they satisfy the following time-dependent minimal conditions
 \begin{eqnarray}
& & \phi^{''} + 3H \phi^{'} + \frac{\partial V(\phi)}{\partial
\phi} = 0 \quad \mbox{or} \quad
 \frac{d}{dt} [ \frac{1}{2}\phi^{'2} + V(\phi) ] + 3H \phi^{'2} = 0
 \end{eqnarray}
 with the Hubble parameter
 \begin{eqnarray}
  H^2 = \frac{1}{3\bar{M}_{Pl}^2} \left( \frac{1}{2}\phi^{'2} +
  V(\phi) + \rho \right)
 \end{eqnarray}
which is obtained from the Einstein equation. Here $H(t) = a'/a$
with $a(t)$ the scale factor in Robertson-Walker model. $\rho$
represents the energy density of radiation or matter, and
satisfies the continuity equation
\begin{eqnarray}
 \frac{d}{dt}\rho + 3H (1 + \omega) \rho = 0
\end{eqnarray}
Here $\omega = p/\rho$ is the ratio of pressure to density.

 We assume that in the first stage of evolution the universe
 is mainly governed by the potential energy characterized by the effective potential $V(\phi)$.
We also assume that the rate $\Gamma$ of particle creation by
$\phi$ in this stage is much smaller than the expansion of
universe, i.e., $\Gamma << 3 H$. On the other hand, it can be
shown that for $\lambda_1 \sim \lambda_2 \sim \lambda_3 \sim
\lambda_4$ and $g_F \sim c_0 \sim 1$, the potential $V_1$ becomes
dominated in the inflationary period. As $V_1(\phi)$ is an
exponential potential $V_1$, the vacuum state is known to get an
exact solution
 \begin{eqnarray}
& & \phi(x) \simeq \phi(t) \simeq \phi_0 - 2M \ln (1 + M_o t),
\quad \lambda_1 g_F^3c_0 e^{\phi_0/M} (8\pi \bar{M}_{Pl})^2 = (3p
-1)p\ M_0^2  \\
& & V(\phi) \simeq  V_1(\phi) = (3- 1/p)\  \bar{M}_{Pl}^2\
\frac{(p M_0)^2}{(1+ M_o t)^2} = (3- 1/p)\ \bar{M}_{Pl}^2 H^2(t)
 \end{eqnarray}
with the power-law index $p$ being determined by the gauge fixing
parameter $g_F$
 \begin{eqnarray}
 p = 16\pi \frac{M^2}{M_{Pl}^2} = 16\pi g_F^2 \left(1 + \frac{3}{4\pi g_F^2}\right)
 \end{eqnarray}
which causes a power-law inflation
\begin{eqnarray}
& & a(t)  = a_0 (1 + M_o t)^{p} \\
& & H(t)  =  \frac{p M_o}{1 + M_o t }
 \end{eqnarray}

 As a predictive case of the theory, taking the Planck mass as the gauge fixing
 scale of the scalar field $\Phi$, we have $g_F = 1$. Thus
 the power-law index is predicted to be
\begin{eqnarray}
 p = 16\pi \frac{M^2}{M_{Pl}^2} = 16\pi \left(1 +
 \frac{3}{4\pi}\right)\simeq 62
 \end{eqnarray}
It is easy to check that for such a large value of $p$ the
effective potential $V(\phi)$ goes via a slowly-rolling down
process. This can be seen from the following condition
 \begin{eqnarray}
 |\phi''|/|3H\phi'| = 1/p \ll 1
 \end{eqnarray}

We can now follow the standard approaches to compute various
cosmological observables by considering the predictive case
$g_F=1$ and $p\simeq 62$. In the leading approximation with
ignoring possible quantum loop corrections of gauge field and
scalar field, the spectral index of the scalar fluctuation in the
power-law inflation is approximately determined by the parameter
$p$
  \begin{eqnarray}
 n_s \simeq 1-\frac{2}{p} \simeq 30/31 \simeq 0.97
 \end{eqnarray}
 The tensor to scalar ratio is also approximately determined by the parameter
 $p$
 \begin{eqnarray}
 r \simeq 16 \left(\frac{\sqrt{3\pi} \phi'}{M_{Pl} H}\right)^2 =
 \frac{16}{p} = 8 (1-n_s) \simeq 0.26
 \end{eqnarray}
 which is consistent with the current cosmological
 observations\cite{TEG}.

 We now consider the constraint from the primordial density fluctuation spectrum
 which is approximately estimated by
  \begin{eqnarray}
      10^{-5}\alt \frac{\delta \rho}{\rho} = \frac{C}{2\pi}
      \frac{H^2}{\phi'} < 10^{-4}
 \end{eqnarray}
 with $C$ the normalization constant $C= 1/2\sqrt{\pi}$.
 It is assumed that the power-law inflation
  is changed when $\Gamma \agt 3H = 3p/t$, i.e., the particle
  creation rate $\Gamma$ due to the variation of $\phi(x)$ becomes larger than
  the expansion rate $H$. Thus the power-law inflation which is supposed starting from the Planck time $t_I = 1/M_o
  =  1/\bar{M}_{Pl} =\bar{t}_P \simeq 2.7 \times 10^{-43} sec $
 will hold until the time $t_{\Gamma} = \frac{3p}{\Gamma}$.
 Combining the above requirement from primordial density fluctuation spectrum, we obtain the following
 constraint for an inflationary epoch
 \begin{eqnarray}
     & &  \frac{1}{4} \left(\frac{p}{\pi}\right)^{3/2} 10^{4} <
      \frac{t_{\Gamma}}{\bar{t}_P} \alt  \frac{1}{4} \left(\frac{p}{\pi}\right)^{3/2}
      10^{5}
 \end{eqnarray}
 Numerically, we have
 \begin{eqnarray}
  2.2 \times 10^{5} <
      \frac{t_{\Gamma}}{\bar{t}_P} \alt  2.2 \times 10^{6}
 \end{eqnarray}
 In this case, the inflationary universe has a size much larger
 than the observable part
 \begin{eqnarray}
   10^{229} <   \frac{a(t_{\Gamma})}{l_0} \alt 10^{291}
 \end{eqnarray}

 The reheating temperature $T_{rh}$ may approximately be estimated at the
 epoch when the total energy density changes from being dominated
 by the effective potential $V(\phi)$ to being radiation
 domination. Namely
 \begin{eqnarray}
 \rho_r = g^{\ast}(T) \frac{\pi^2}{30} T^4 \simeq 3
 \frac{M_{Pl}^2}{32\pi t^2}
 \end{eqnarray}
 with $g^{\ast}(T)$ is the effective number of helicity states and
 is of order $10^2$. Suppose that the reheating process takes place rapidly, namely $t_{rh} \sim t_{\Gamma}$,
 the reheating temperature is found to be
 \begin{eqnarray}
  2.6 \times 10^{14}\ \mbox{GeV} \alt T_{rh} < 0.85 \times 10^{15}\
  \mbox{GeV}
 \end{eqnarray}
 which is not much lower than the scale of grand unification.

 With the above constraint, the horizon
  problem is well solved since it only needs to satisfy the following condition
  \begin{eqnarray}
     \left( \frac{a(t_{\Gamma})}{a_0}\right)^{(p - 1)/p} \simeq
   \left( \frac{t_{\Gamma}}{\bar{t}_P}\right)^{p -1}     > 5\times 10^{29}
   \frac{T_{rh}}{\bar{M}_{Pl}} \agt 5.3\times 10^{25}
 \end{eqnarray}
 a numerical low bound for $t_{\Gamma}$ is found to be
  \begin{eqnarray}
      \frac{t_{\Gamma}}{\bar{t}_P} \agt 2.6
 \end{eqnarray}
 which is much weaker than the one from the primordial density fluctuation
 spectrum.

 It is seen from the above demonstration that the power-law inflation
 with $g_F=1$ and $p\simeq 62$ remains consistent with the current
 cosmological data though it seems only marginally allowed. For the gauge fixing scale
 to be slightly higher than the Planck scale, for instance $g_F = \sqrt{2}$, one has $ p \simeq 113$.
 Repeating the same computations, we are led to the
 following results
 \begin{eqnarray}
 n_s \simeq 1-\frac{2}{p} \simeq 0.98, \qquad r \simeq
 \frac{16}{p} \simeq 0.14
 \end{eqnarray}
 which is firmly in the allowed region from the current cosmological observations\cite{TEG}.

 Note that in computing the time period $t_{\Gamma}$ of inflation,
 it is assumed that a rapid and successful reheating process takes place
 after the inflation driven by the potential energy $V(\phi)$,
 and the created particles are to be thermalized in a short time ($\ll 1/H(t_{\Gamma})$).
 The power-law inflation is in general changed due to the
 variation of the field $\phi$ via the particle creation.
 Therefore, around the epoch when $\Gamma \sim 3 H$, one may
  solve two correlated continuity equations
 \begin{eqnarray}
& & \frac{d}{dt} [ \frac{1}{2}\phi^{'2} + V(\phi) ] + 3H \phi^{'2}
= -\Gamma \Delta \\
& & \frac{d}{dt}\rho_r  + 4H \rho_r = \Gamma \Delta
\end{eqnarray}
with $\Delta = \phi'^2$ \cite{ASTW,MST}. The magnitude of $\Gamma$
relies on the fields which interact with the scalar field $\phi$
and the coupling strength among them. In general, one also needs
to consider the effect of quantum gauge field $a_{\mu}$ in the
reheating process. Here we shall not discuss in detail for the
process of reheating, which is beyond our aim in the present
paper.

 In conclusion, we have shown that based on the conformal scaling
gauge symmetry in the presence of gravity, there must exist a
conformal scalar field $\Phi(x)$ and a gauge field $A_{\mu}$. A
gauge invariant potential action can in general be constructed by
adopting a real Wilson-like line element of gauge field and the
scalar field. The conformal scaling gauge symmetry has explicitly
been broken down by fixing the gauge to match the Einstein-Hilbert
action for gravity. A nontrivial background field solution of pure
gauge has been found to result in an exponential potential model
of scalar field, which leads to a power-law inflationary universe.
The power-law index $p$ is determined by the gauge fixing
parameter $g_F$ which is a unit when the gauge fixing scale is
taken to be the Planck mass. The model is consistent with the
current cosmological observations for a natural large value of the
power-law index $p \simeq 62\sim 113$ which is corresponding to
the gauge fixing parameter $g_F =1\sim \sqrt{2}$. In this case,
the spectral index of the scalar fluctuation and the tensor to
scalar ratio are predicted to be $n_s \simeq 1-2/p \simeq 0.97\sim
0.98$ and $r \simeq 16/p \simeq 0.26\sim 0.14$ respectively. As
the potential energy plays a role of dark energy, the ratio of
pressure to energy is estimated to be $\omega \simeq - n_s \simeq
-0.97\sim - 0.98$. The expected reheating temperature $T_{rh}$ is
below the grand unification scale and has a typical value: $T_{rh}
\simeq 10^{14}$ GeV. It is noted that in this paper we have only
considered the simple cases at leading approximation and ignored
quantum corrections of gauge field as well as other possible
interactions. Some important issues such as the ending of
inflation and the process of reheating have not been studied as
they are beyond our purposes in the present paper. Also the
question on the unexpectedly small anisotropy of CMB at large
angles\cite{CC} needs to be understood. Nevertheless, it remains
unclear whether it is a real anomaly or it is just a manifestation
of cosmic variance\cite{GE}.

Acknowledgement: Author would like to thank R.G. Cai, M. Li, J.X.
Lu, X.C. Song, Y.Z. Zhang and many colleagues for useful
discussions and conversations. This work was supported in part by
the projects of NSFC and CAS, China.

\end{document}